\documentclass[aps,prl,reprint,groupedaddress]{revtex4-1}
\usepackage{graphicx} % Include figure files
\usepackage{dcolumn} % Align table columns on decimal point
\usepackage{bm}% bold math
\usepackage{hyperref}% add hypertext capabilities
\usepackage{siunitx}
\usepackage{float}
\usepackage{graphicx} %use graph format
\usepackage{epstopdf}
\usepackage{verbatim}
\usepackage{array}
\usepackage{booktabs}
\usepackage{multirow}
\usepackage{amsmath}
\usepackage{amssymb}
\usepackage{setspace}

\begin{document}

\title{A novel approach to light cluster production in heavy-ion collisions}
\author{Hui-Gan Cheng }
\author{Zhao-Qing Feng }
\email{Corresponding author: fengzhq@scut.edu.cn}

\affiliation{School of Physics and Optoelectronics, South China University of Technology, Guangzhou 510640, China}

\date{\today}

\begin{abstract}
 The issue of cluster production in heavy-ion collisions is addressed in a new manner, by implementing cluster correlation into the quantum molecular dynamics (QMD) transport model. We demonstrate for the first time, the good potentialities of this popular transport approach in the description of light cluster production including the deuteron, triton, $^{3}$He and $\alpha$ particle in heavy-ion collisions at intermediate energies. Both the INDRA and FOPI experimental data of the total multiplicities of light clusters and the charge distributions of heavier fragments are reasonably reproduced by the unified approach. The effects of both the cluster binding energies and the pauli repulsion are also shown to play crucial roles in the production of clusters.

\begin{description}
\item[PACS number(s)]
21.80.+a, 25.70.Pq, 25.75.-q             
\end{description}
\end{abstract}

\maketitle

%Introduction
Renewed attention has been attracted to the study of clustering in nuclear physics in past decades. The emergence of clusters is of nontrivial interest, sometimes central importance, in the study of nuclear physics\cite{BB10,BB12,BB14, Ho12,Eb12,Re18,Br08,Ro15}. In the particular study of Heavy-Ion Collisions (HICs) at incident energies of tens of MeV to hundreds of MeV, for instance, the yields of deuterons, tritons, $^{3}$He and $\alpha$ particles are comparable with or even greater than that of protons\cite{On19}. The huge effects of clusters on the equation of state (EoS) of nuclear matter has already been revealed in the low-density regime\cite{Na10}. Thus the crucial role played by clusters in the dynamical evolution of the reaction system is also supposed to have an impact\cite{On19,Ik16} on the final-state observables which furnish probes to the nuclear EoS in the high-density regime\cite{Li02,Sp21,Da02}. Therefore cluster is an unavoidable aspect which deserves explicit treatments in transport approaches for HIC. In astrophysical studies, like pasta phases\cite{Ca17,Ho15,Li20}, the rich cluster contents in dilute and warm nuclear matter\cite{Qi12,Ha12,Pa20} also has significant relevance in the core-collapse of supernova and the properties and evolution of compact stars\cite{Su08,Oe17}.

In HICs at higher energies, for example, several GeV/nucleon in the hypernuclear experiments at FAIR\cite{Gu06, Ay16, Ge03} in Germany, RHIC-STAR\cite{Ab22, Ad20, Ab23} in the United States, NICA in Russia\cite{Nica}, and HIAF\cite{Ya13} in China in the near future, hypernuclei might be created. They furnish unique laboratory to study two-body hyperon-nucleon and three-body hyperon-nucleon-nucleon interactions which lie at the heart of the understanding on the inner structure of compact stars\cite{Ge20,Lo15}.  The dynamics of hypernucleus production in the mid-rapidity regions may be reasonably described by the mechanism of baryon coalescence in central HICs\cite{Ab23}. However, in central collisions, only clusters of $A=2,3$ are richly produced\cite{Re10} in the violent fireball as understood by the mechanism of coalescence, and this is even more the case at still higher energies\cite{Ai20}. On the other hand, for light- or medium-mass reaction systems, $^{6}$Li+$^{12}$C\cite{Fe20,Ra15,Fe19}, $^{20}$Ne+$^{12}$C\cite{Fe20} and $^{40}$Ca+$^{40}$Ca\cite{Ch22} for examples, the majority of the light hypernucleus yields is centered at the spectator rapidity region. Also in peripheral collisions with heavy reaction systems, hyperclusters are copiously produced in fireball-spectator fragmentation reactions\cite{Bo13,Bo17}.Thus a more interesting or perhaps more copious source of hypernuclei can be envisioned if we consider the capture of hyperons by the lighter clusters produced in the multi-fragmentation of the less violent mixed region between the fireball and the spectators, or of the hot spectators themselves\cite{Bo11}. And in this, it is thus a prerequisite to properly account for the issue of cluster production in the fragmenting spectator and mixed regions which break up at intermediate energies. This may also open the perspective to study the liquid-gas phase transition of nuclear matter with hyperons and the production of hypernuclei with extreme $N/Z$ ratios.

In this letter, we address the issue of clusters in heavy-ion collisions from a new perspective, by implementing into the QMD model the kinetic production of clusters including deuterons, tritons, $^{3}$He and the alpha particle for the first time, achieving a good description of the production of both light clusters and heavier fragments in a unified manner. The effects of both the cluster binding energies and the pauli repulsion on clusters \cite{Xu16} are also shown to play vital roles in the production of clusters.

In the past and in very recent years, the kinetic approach of cluster production has well been incorporated into transport models like the Anti-symmetrized Molecular Dynamics model (AMD) \cite{On13,On16} or the Boltzmann-Uehling-Uhlenbeck (BUU) \cite{Da91,Da92,Ro88,Ku01,St21,Su22,Wa23} which are capable of describing the production of clusters even up to the alpha particle\cite{On13,On16,Wa23}. Much as they promise the excellent theoretical tools in the study into the effects of clusters on the many facets of HICs, for example the extraction of the EoS at high density\cite{Ik16}, fragment formation\cite{Ti18,Fr23}, the suppression of triton production in ultra-relativistic HICs\cite{Su22}, and so on, it is rather intriguing, however, to also see similar attempts based on another popular class of models, the QMD-type transport model\cite{Ai91}. On one hand, it saves computational resources drastically, and on the other hand, it accounts for the branching of the reaction system into final states of different fragment partitions. Therefore, it furnishes the excellent opportunity to investigate into the above mentioned cases where both cluster and strangeness join to play an interesting role. Thus the extension to light cluster production up to $\alpha$ based on QMD in multi-fragmentation reactions would be of nontrivial progress. In this, only very recently the status quo has begun to come to light, with a partial step made to the Parton-Hadron-Quantum-Molecular Dynamics model (PHQMD) \cite{Co23} by incorporating the production of the deuteron as an extra degree of freedom in the study of central HIC at highly or ultra-relativistic energies. Still, to study the production of light clusters in the break-up of the fireball and the hot spectators requires the extension up to the alpha cluster and the entire charge spectrum, and it thus would be entertaining to see a possible advance.

In this work, we treat the light clusters as composite particles composed of nucleons, the same strategy as adopted in the model of AMD-cluster. The fermionic nature of the nucleon is partially taken into account by considering a continuous phase-space constraint which acts like a pauli potential. A reasonable reproduction of the INDRA \cite{Hu03,Zb07} and FOPI \cite{Re10} experimental yields of light clusters and fragment charge distributions is obtained in a unified manner, furnishing a possible starting point of the extension to the very applications we mentioned above.
%Model description
In the treatment of cluster production, we follow the method proposed by Ono in AMD-cluster, as documented in previous articles\cite{On13,On16}. The basic idea is to include quantum transition to clustered states up to the production of $\alpha$ as possible final states of the scattering between two nucleons $\mathrm{N}_{1}\mathrm{N}_{2}$. Starting from the non-clustered $N\!N$ scattering final state, repeated steps are taken by constructing the projection operator $\hat{P}=\sum_{ij} |P_{i}\rangle (N^{-1})_{ij} \langle P_{i}| $ for the subspace of clustered states at each step, to consider all scattering channels $C_{1}+C_{2}\rightarrow C_{3}+C_{4}$ between nucleon-nucleon, nucleon-cluster and cluster-cluster in a unified manner. The actual treatment, however, is much more complicated. Denote the state of the reaction system before $N\!N$ scattering as $\left | O  \right \rangle$, and denote the state after ordinary $N\!N$ scattering without considering clusters and energy conservation as $\left | P  \right \rangle$. We need to adjust the relative momentum between $\mathrm{N}_{1}$ and $\mathrm{N}_{2}$ to a state $\left | Q  \right \rangle$ from which the energy-conserving final state can be constructed. If we assume the transition amplitude of the process to be $P(C_{1}+C_{2} \rightarrow C_{3}+C_{4}){|T(\tilde{p}_{\mathrm{rel}})|}^{2}$ where $|T(\tilde{p}_{\mathrm{rel}})|$ is the ordinary $N\!N$ scattering transition amplitude evaluated at the average relative $N\!N$ momentum $\tilde{p}_{\mathrm{rel}}$ of $\left | O  \right \rangle$ and $\left | Q  \right \rangle$, the differential cross-section of the process can be cast as follows
\begin{eqnarray}
\frac{d \sigma}{d \boldsymbol{\Omega}}=&&P(C_{1}+C_{2} \rightarrow C_{3}+C_{4}) \times
	\nonumber \\
 &&\frac{v_{\tilde{p}_{\mathrm{rel}}}}{v} \frac{\left|[\partial e(k) / \partial k]_{k=\tilde{p}_{\mathrm{rel}} \mid}\right|}{\left|\left[\partial H\left(p_{f}\right) / \partial p_{f}\right]_{p_{f}=p_{\mathrm{rel}}}\right|} \frac{p_{\mathrm{rel}}^{2}}{\tilde{p}_{\mathrm{rel}}^{2}}\left[\frac{d \sigma_{\mathrm{NN}}}{d \boldsymbol{\Omega}}\right]_{\tilde{p}_{\mathrm{rel}}}.
\end{eqnarray}
Here $v$ is the $N\!N$ relative velocity in $\left | O  \right \rangle$ and $v_{\tilde{p}_{\mathrm{rel}}}$ is that under $\tilde{p}_{\mathrm{rel}}$. $p_{\mathrm{rel}}$ is the relative momentum between $\mathrm{N}_{1}\mathrm{N}_{2}$ in $\left | Q  \right \rangle$. $e(k)$ is the kinetic energy of the two nucleons in free space and $H$ is the total energy of the reaction system. Both  $e(k)$ and $H(p_f)$ are functions of the relative momentum between $\mathrm{N}_{1}\mathrm{N}_{2}$. The last factor in the expression is the ordinary free-space differential $N\!N$ cross-section evaluated at $\tilde{p}_{\mathrm{rel}}$.
\begin{eqnarray}
	H&&=\sum_{i}\frac{{\mathbf{p} _{i}}^{2}}{2m}+\frac{\alpha}{2}
	\sum_{\substack{i,j \\ j \neq i}} \frac{\rho_{i j}}{\rho_{0}}+\frac{\beta}{1+\gamma} \sum_{i}\left(\sum_{\substack{j, j \neq i}} \frac{\rho_{i j}}{\rho_{0}}\right)^{\gamma}
	\nonumber\\
	&&+\frac{C_{\mathrm{sym}}}{2} \sum_{\substack{i,j \\ j \neq i}} t_{z_{i}} t_{z_{j}} \frac{\rho_{i j}}{\rho_{0}}
	+\frac{\mathrm{g}_{\text {sur }}}{2}
	{\sum_{\substack{i,j \\ j \neq i}}}^{'}\left[\frac{3}{2 L}-\left(\frac{\mathbf{r}_{i}-\mathbf{r}_{j}}{2 L}\right)^{2}\right] \frac{\rho_{i j}}{\rho_{0}}
	\nonumber\\
	&&+\sum_{i}^{N_{C} } E_{\mathrm{z.p.} }^{i}+\sum_{i}^{N_d} V_{\mathrm{corr} }e^{-{r_i}^{2}/4L}+V_{\mathrm{Coul}}
\end{eqnarray}
For the treatment of the mean-field evolution of the reaction system, we employ the usual standard form of the mean-field Hamiltonian in the LQMD model \cite{Fe11}, but with some slight modifications as in the above equation. We turn off the the surface interaction term between any two nucleons among each cluster, as indicated by the extra prime added to the summation in the $\mathrm{g}_{\mathrm{sur} }$ term, so that all the nucleons within a cluster move as a single object under the surface potential. We emphasize that the surface term which is abandoned in, for example, the PHQMD model\cite{Co23}, is a vital ingredient for the emission of the formed clusters. Without this term, clusters are all glued to each other or to bigger fragments and not a single cluster can thus be set free to the gas phase in QMD. The last but two term is a summation over the zero-point potential energy among each cluster. This represents the internal quantum kinetic energy of nucleons within a cluster, and we adopted the form defined in ref.\cite{Ma96}. But for the width and the smearing parameters, we take $a=0$ fm and $b=2.25\times2.25$ fm$^{2}$.We note in passing that the detailed shape of the zero-point potential has only negligible effects on the production of clusters and heavier fragments, as we have verified. We choose the above parameters just to guarantee that no local energy-minima are encountered in the dissolution of clusters, as will be described later. Finally, in the last but one term, the binding energy of the deuteron is corrected by artificially adding a term, where $V_{\mathrm{corr} }=1 \ \mathrm{MeV}$, $r_i$ is the relative distance between the wave-packet centers of the two nucleons, and $L=1.75$ fm$^{2}$ is the wave-packet width parameter tuned here for LQMD to reasonably reproduce the binding energy of all light clusters. With the above prescription, the binding energy per nucleon of deuteron is then 1.15 MeV, close to the experimental value. Other parameters in the above equation are $\alpha=-226.5\ \mathrm{MeV}$, $\beta=173.7\ \mathrm{MeV} $, $\gamma=1.309$, $C_{\mathrm{sym}}=38\ \mathrm{MeV}$, $\mathrm{g}_{\mathrm{sur}}=23\ \mathrm{MeV\cdot fm^{2}} $, $\rho_{0}=0.16\ \mathrm{fm}^{-3}$.

In AMD-cluster, cluster correlation is only allowed where the nucleon density is above $\rho_{cut}=0.125$  fm$^{-3}$\cite{OnTk}. In AMD or the Extended Quantum Molecular Dynamics(EQMD)\cite{Ma96}, a very large portion of the nucleon kinetic energy appears as the quantum zero-point motion of the Gaussian wave function and the intra-nuclear motion of the nucleons in the initial state are completely frozen in these models. For this reason, the centroids of the nucleons' wave-packets move faster in QMD than in AMD or EQMD so that it is more difficult to form clusters due to lower overlap between the wave-packets. Meanwhile the density distribution of the system fluctuates more violently with faster moving nucleons. Thus, with the same form of in-medium $N\!N$ cross-section employed\cite{On16}, to ensure that the same amount of cluster correlation is seen in both AMD-cluster and QMD, we are forced to adopt a higher density cut $\rho_{cut}=0.170$  fm$^{-3}$ in our case. This may be avoided alternatively by artificially enhancing the nucleon-nucleon (NN) scattering cross-section or the cluster formation probabilities.

The fermionic nature of the nucleon is a fundamental aspect in the formation and evolution of clusters in HIC. That the nucleon is a fermion is important in two aspects mainly due to the exchange terms in the mean-field Hamiltonian of a fermionic system. The exchange part of the kinetic energy term leads to the pauli repulsion between fermions which can be mimicked by introducing a pauli potential\cite{Bo88,Pe91,Pe92} or with the method of phase-space constraint\cite{Pa01} in QMD. The exchange part in the interaction term accounts for the change of the strength of binding between the nucleons within a cluster moving in nuclear media. These jointly lead to the Mott effect, the dissolution of clusters in nuclear media.

The method of phase-space constraint was proposed by Papa et al\cite{Pa01} in their Constrained Molecular Dynamics model (CoMD) to render $\textit{f}_{i}$'s below 1 for all nucleons in the course of mean-field evolution by a series of NN scatterings, but this is not viable when it comes to the case with clusters, since the evolution of nucleons is not continuous in momentum space with this method, which would destroy all formed cluster structures. For a remedy to this deficiency, a more delicate treatment is in order. We define a phase-space compactness $U_{i}$ in the form of a pauli potential defined in Ref.\cite{Ma96}, to measure the compactness of nucleons in each phase-space region $P_{i}$ defined in the following, and lower $U_{i}$ continuously by the technique of frictional cooling\cite{Ka95} until the $\textit{f}_{i}$'s are all below a given bound, so that this method acts like a pauli potential. The $P_{i}$'s are identified by an elaborate Minimum Spanning Tree(MST) procedure which divides the system into different phase-space regions, for each spin-isospin separately. We require that the $k_{ij}\le 6 a \sqrt{L}$ is fulfilled between each pair of nucleons $i$ and $j$ within an identified phase-space region, where $k_{ij}^{2}=(a(\textbf{r}_{i}-\textbf{r}_{j}))^{2}+((\textbf{p}_{i}-\textbf{p}_{j})/2a \hbar)^{2}$ with $a=0.4$ fm$^{-1}$.
The treatment entails the following details. First, a usual MST procedure with $r_{0}=3.5$ fm and $p_{0}=200$ MeV/c\cite{Ch21} is applied to the whole system, and for each nucleon $i$, the size of the MST fragment that it resides is defined as $NF_{i}$.  We further classify the phase-space regions into those composed of light fragments and those of heavy fragments. If over 85\% of the constituent nucleons are of $2\le NF_{i}\le 20$, a phase-space region is considered as composed of light fragments. Otherwise it is composed of heavy fragments. For each nucleon $i$, define a quantity $M_{i}$ which we may call 'marginality' measuring the closeness of $i$ to the boundary of the nucleon distribution in space. $M_{i}$ is the number of nucleons within a sphere of radius $3 \sqrt{L}$ and centered at $i$. A phase-space region is regarded as belonging to dilute area, if over 83\% of its constituent nucleons are of $M_{i}\le 12$. Then the bound upon the phase-space occupation of each nucleon $i$ is defined as a hyperbolic tangential function,
\begin{gather}
	f_{boun,i}=f_{boun,l}+
	\frac{f_{boun,u}-f_{boun,l}}{2} (\text{tanh}\frac{M_{i}-\frac{M_{u}+M_{l}}{2}}{\frac{M_{u}-M_{l}}{4}}+1),
\end{gather}
where the upper and the lower values of the bound are $f_{boun,u}=0.95$ and $f_{boun,l}=0.65$, and the corresponding marginalities $M_{i}$ are $M_{u}=17$ and $M_{l}=8$. We further define similarly a 'tolerance' for each nucleon,
\begin{gather}
	\epsilon_{boun,i}=\epsilon_{boun,l}-\frac{\epsilon_{boun,l}-\epsilon_{boun,u}}{2} (\text{tanh}\frac{M_{i}-\frac{M_{u}+M_{l}}{2}}{\frac{M_{u}-M_{l}}{4}}+1),
\end{gather}
where $\epsilon_{boun,l} = 0.10$ and $\epsilon_{boun,u} = 0.05$. These two functions $f_{boun,i}$ and $\epsilon_{boun,i}$ come to an approximate lower platform at $M_{l}$ with $f_{boun,i}\approx f_{boun,l}=0.65$ and $\epsilon_{boun,i}\approx \epsilon_{boun,l}=0.1$, whereas they come to an upper platform at $M_{u}$ with $f_{boun,i}\approx f_{boun,u}=0.95$ and $\epsilon_{boun,i}\approx \epsilon_{boun,u}=0.05$.

With the above definition, the procedure of phase-space constraint is carried out as follows. Each time, a step of constraint is separately imposed on all the phase-space regions for each spin-isospin except for those for which $f_{i}< f_{boun,i}$ is satisfied for all the constituent nucleons. Within a time step of dynamical evolution of the reaction system, repeated steps are performed for the whole reaction system until $\epsilon_{1}< \epsilon_{2}$, where $\epsilon_{1}$ is the average value taken of $f_{i}-f_{boun,i}$ for those nucleons involved in the constraint and for which $f_{i}>f_{boun,i}$ still after the step. $\epsilon_{2}$ is the average value taken of $\epsilon_{boun,i}$ for those nucleons with which $\epsilon_{1}$ is calculated. And for each new step, the MST division of the whole reaction system into phase-space regions is renewed based on a different ordering of the nucleons so that the constraint is a stochastic but continuous process. The frictional cooling equation as described in Ref.\cite{Ka95} is solved with the the frictional coefficients, $\lambda=-4$ and $\mu=-4$, and $dt = 0.1$ fm/c, a same size of time step adopted for the dynamical evolution of the reaction system. Meanwhile, in the frictional cooling process all 16 constraints including the total energy, the total momentum, the center of mass, the total angular momentum, and the six components of the quadrapole tensor $\sum_{i} r_{\alpha i} r_{\beta i}$, $\alpha,\beta=1,2,3$ of the whole reaction system are kept constant. In this way, the process of phase-space constraint acts like that of a pauli potential and the formation of cluster structures in dilute regions at the disintegration of the system can thus be favored by imposing a tighter phase-space constraint, as indicated by $f_{i}\le f_{boun,i}$ which has a tightest bound, $f_{boun,l}=0.65$ in the most dilute areas. The constraint on the quadrapole tensor is to render the lengths of the main axes of the system fixed, to avoid evolution into bizzare shapes, and to dictate that the constraint acts more on momentum space. But for the phase-space regions belonging to dilute areas which are usually dominated by exotic structures, this constraint is not imposed and only 10 constraints are imposed. When the evolution of the reaction system is running normally after the end of the most violent stages, usually one step of phase-space constraint is sufficient within each time step of the dynamical evolution of the system.

In CoMD, an essential step is to initialize the ground state of a nucleus by lowering to the energy minimum under phase-space constraint. But here we can by no means do so since when the nucleus is very close to its ground state and thus nucleons are more and more compact in phase space, a global reorganization of nucleons in phase space is demanded, and this can only be achieved when a discontinuous phase-space evolution as in CoMD is allowed.

For the dissolution of clusters in mean-field evolution, at the end of each time step of mean-field evolution, a MST procedure is applied among the nucleons within each existing clusters, in which a MST criterion $k_{ij}\le 1.75$ is satisfied between any two nucleons in a MST fragment. If a cluster is identified as consisting of more than one MST fragments, the mother cluster is destroyed and the daughter clusters are registered. For the formation of bigger clusters, a MST procedure is applied to the entire system with $r_{0}=3.5$ fm and $p_{0}=200$ MeV/c as mentioned earlier. If a MST fragment($2\le A\le 4$) contains, as its constituents, smaller existing clusters, all the smaller clusters are destroyed and the MST fragment is registered as a new existing cluster. In the above processes, the conservation of energy is treated via frictional cooling.

%Results
%%%%%%%%%%%%%%%%%%%%%%%%%%%%%%%%%%%% figure 1 %%%%%%%%%%%%%%%%%%%%%%
\begin{figure}
	\includegraphics[width=9.3 cm]{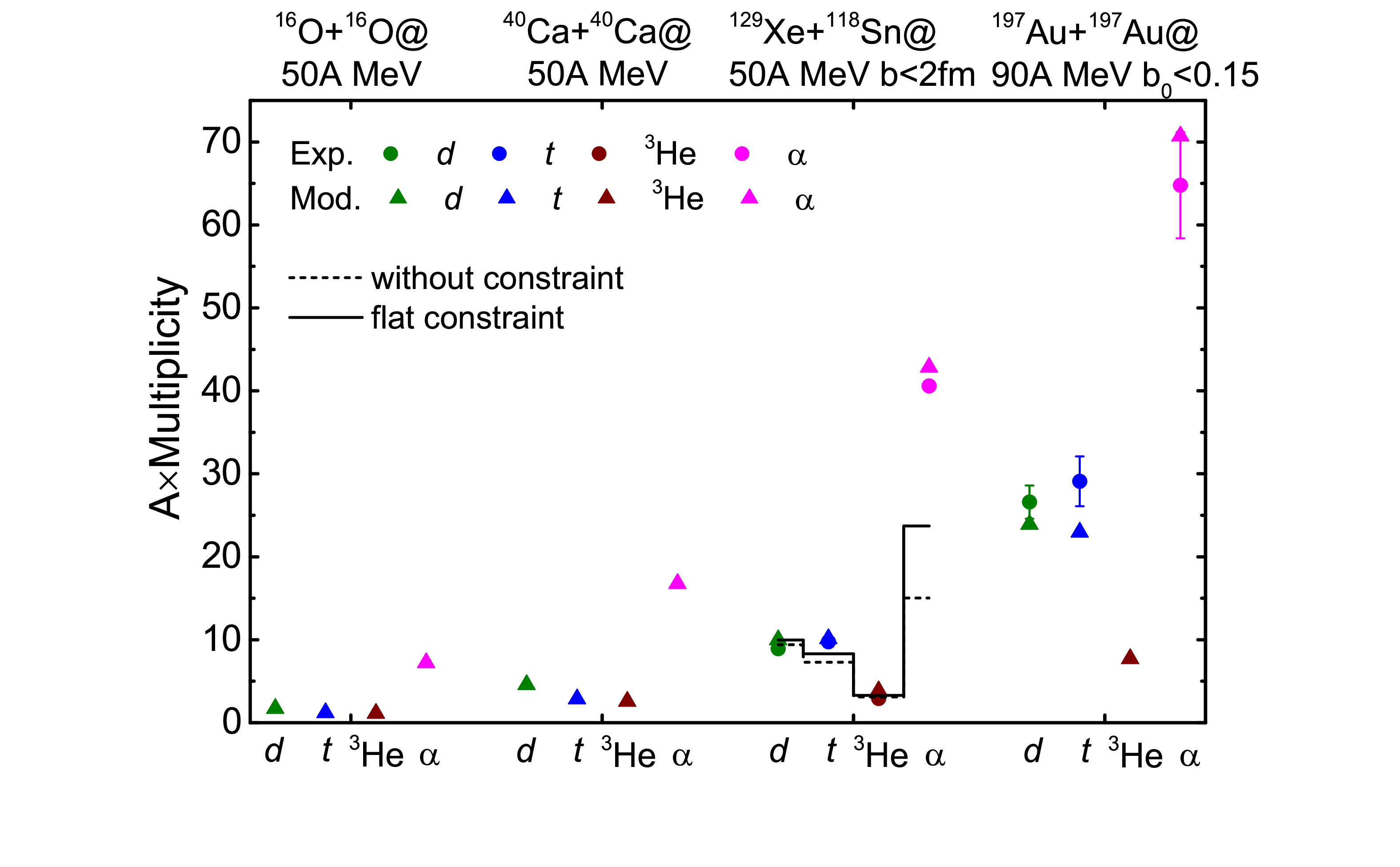}
	\caption{The numbers of nucleons of each type of clusters including the deuteron(olive), triton(blue), $^{3}$He(wine) and $\alpha$ particle(magenta), are plotted for the results of the model(solid triangles) and that of experiments(solid circles), with the INDRA data\cite{Hu03} for $^{129}$Xe+$^{118}$Sn@50A MeV and FOPI data\cite{Re10} for $^{197}$Au+$^{197}$Au@90A MeV, respectively. The data of $\alpha$ in the latter reaction system is taken from that of $^{197}$Au+$^{197}$Au@120A MeV, as explained in the text. The results without phase-space constraint(short-dashed line) and with 'flat' phase-space constraint(solid line) are also plotted for $^{129}$Xe+$^{118}$Sn@50A MeV. }
\end{figure}
%%%%%%%%%%%%%%%%%%%%%%%%%%%%%%%%%%%%%%%%%%%%%%%%%%%%%%%%%%%%%%%

%here!!!

In Fig. 1, we present the numbers of nucleons contained in the final-state deuterons, tritons, $^{3}$He and $\alpha$ emitted in central HIC with reaction systems of different mass numbers and different isospin asymmetries, at incident energies of several tens of MeV. Triangles stand for results of the model, and solid circle for that of experiments. For the last two reaction systems on the right hand side, experimental data are displayed. For $^{129}$Xe+$^{118}$Sn@50A MeV, the INDRA data\cite{Hu03} are plotted, and for $^{197}$Au+$^{197}$Au@90A MeV, the FOPI data\cite{Re10}. Since in the FOPI data, the multiplicity of $\alpha$ and $^{3}$He are not explicitly given for $^{197}$Au+$^{197}$Au@90A MeV, we substitute the $\alpha$ multiplicity of $^{197}$Au+$^{197}$Au@120A MeV, in that the experimental $\alpha$ yield is a constant within error bars in the incident energy range around 120\emph{A} MeV. Hot fragments are all de-excited at 600 fm/c through the GEMINI code. For all the reaction systems shown here, the yields of $\alpha$ prevail. And for the reaction systems for which the experimental results are available, the results of our model agrees with the data reasonably. On examining the figure more carefully, it can also be pointed out that the isospin asymmetries of the reaction systems are also correctly reflected on the relative yields of triton and $^{3}$He. For $^{16}$O+$^{16}$O and $^{40}$Ca+$^{40}$Ca which are isospin-symmetric, the yield ratio triton/$^{3}$He is almost 1, whereas in $^{129}$Xe+$^{118}$Sn and $^{197}$Au+$^{197}$Au which are neutron-rich, this ratio is greater than 1. When the phase-space constraint is not applied, as indicated by the dark short-dashed line for $^{129}$Xe+$^{118}$Sn, the yield of $\alpha$ is almost the same as that of triton+$^{3}$He, and lower than that of deutrons. When this method is turned on but with a same $f_{boun,i}$, that is, $f_{boun,i}=0.95$ and $\epsilon_{boun,i}=0.05$ for all nucleons, as indicated by the dark solid line dubbed as 'flat constraint', the yield of $\alpha$ almost doubles. And when we apply stronger constraint to dilute regions, as described in the context earlier, the yield of $\alpha$ doubles again to match the experimental data. This can be attributed to the both the global improvement of phase-space distribution of nucleons over the entire system, and to a complex of factors caused by the stronger phase-space constraint in dilute regions. These mainly include the aggregation of smaller clusters and nucleons into bigger clusters, the reduced pauli-blocking in forming clusters through scattering, the increased outward pauli repulsion of the boundary regions upon $\alpha$\cite{Xu16}, and finally the interplay among all these effects, etc. But it is difficult to disentangle these interweaving effects from one another anyway. For this reaction system, the time evolution of all gas-phase clusters is also plotted in Fig. 2. We observe that above 120 fm/c, the multiplicities of all clusters already reach a approximate platform so that as far as the production of clusters is concerned , the time cut to switch from the dynamical stage to the stage of statistical decay is not so important in our case.
%%%%%%%%%%%%%%%%%%%%%%%%%%%%%%%%%%%% figure 2 %%%%%%%%%%%%%%%%%%%%%%
\begin{figure}
	\includegraphics[width=8.7 cm]{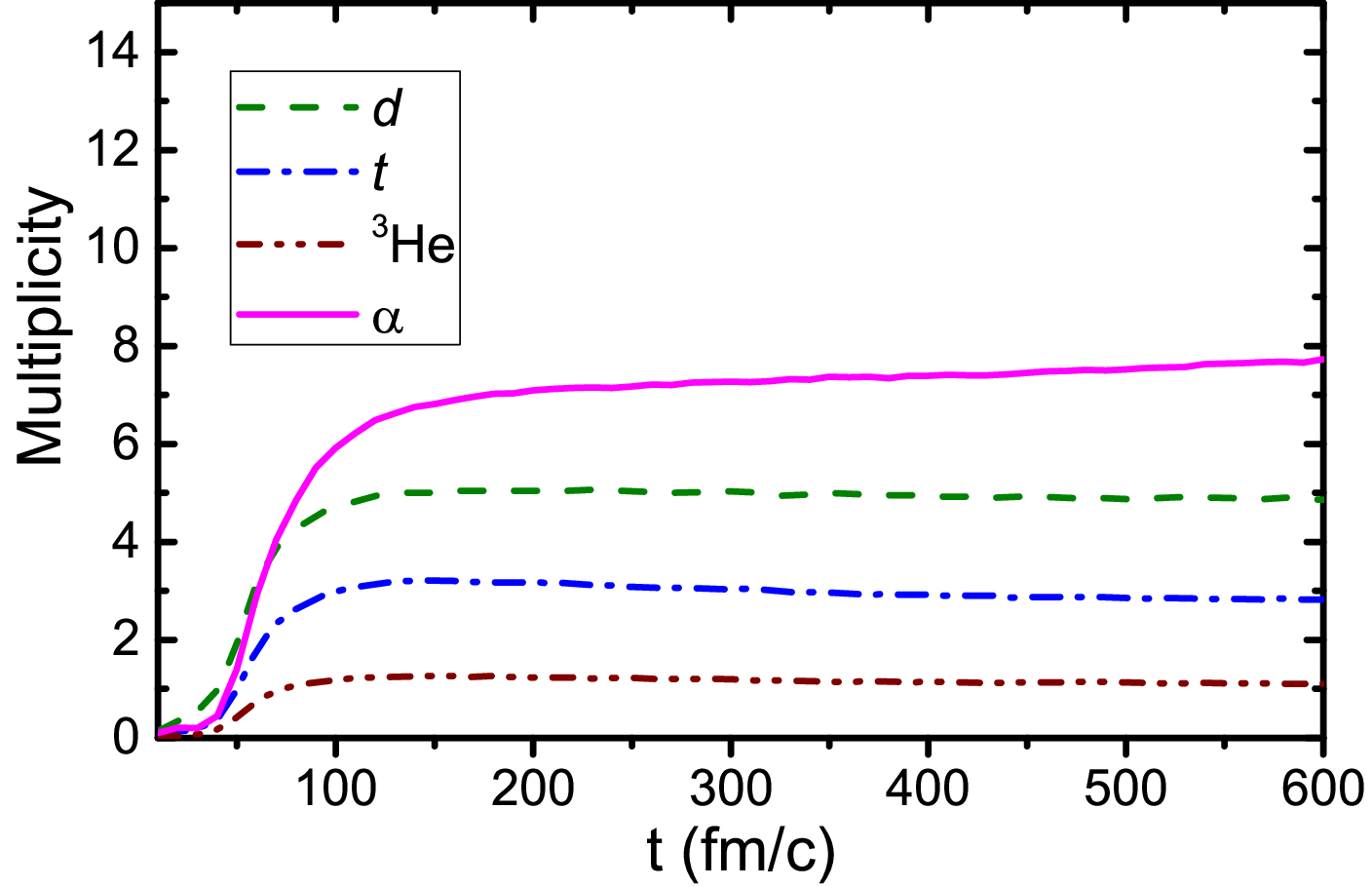}
	\caption{The multiplicities of gas-phase clusters including the deuteron(dashed line), triton(dot-dashed line), $^{3}$He(dot-dot-dashed line) and the $\alpha$ particle(solid line) are plotted as functions of the evolution time of the reaction system $^{129}$Xe+$^{118}$Sn@50A MeV. Here the time starts at 10 fm/c when the two nuclei already overlap.}
\end{figure}
%%%%%%%%%%%%%%%%%%%%%%%%%%%%%%%%%%%%%%%%%%%%%%%%%%%%%%%%%%%%%%%

In AMD-cluster, the copious amount of free $\alpha$ probably mainly emerge as outgoing nucleons or primordial clusters of $A=2,3$, capturing the neighboring partners to form $A=4$ through fermionic mean-field evolution, in their way to be released. In our case, unfortunately, we found that this effect of aggregation is still not strong enough even with phase-space constraint, and thus a part of this process is alternatively and effectively mimicked by turning these lighter objects into $\alpha$ through their scatterings in the expanding fireball. This is achieved by an adjustment of the phase-space distribution of nucleons as outlined in the model description. So the inner-working of our model in the description of cluster correlation is bound to differ from that of AMD-cluster on this point. This  resembles in a sense the treatment in BUU, where all possible(in our case part of) sources of cluster formation are effectively realized in the form of microscopic scatterings. In this, a future improvement towards a more realistic description of $\alpha$ formation is called for.
%%%%%%%%%%%%%%%%%%%%%%%%%%%%%%%%%%%% figure 3 %%%%%%%%%%%%%%%%%%%%%%
\begin{figure*}
	\includegraphics[width=16 cm]{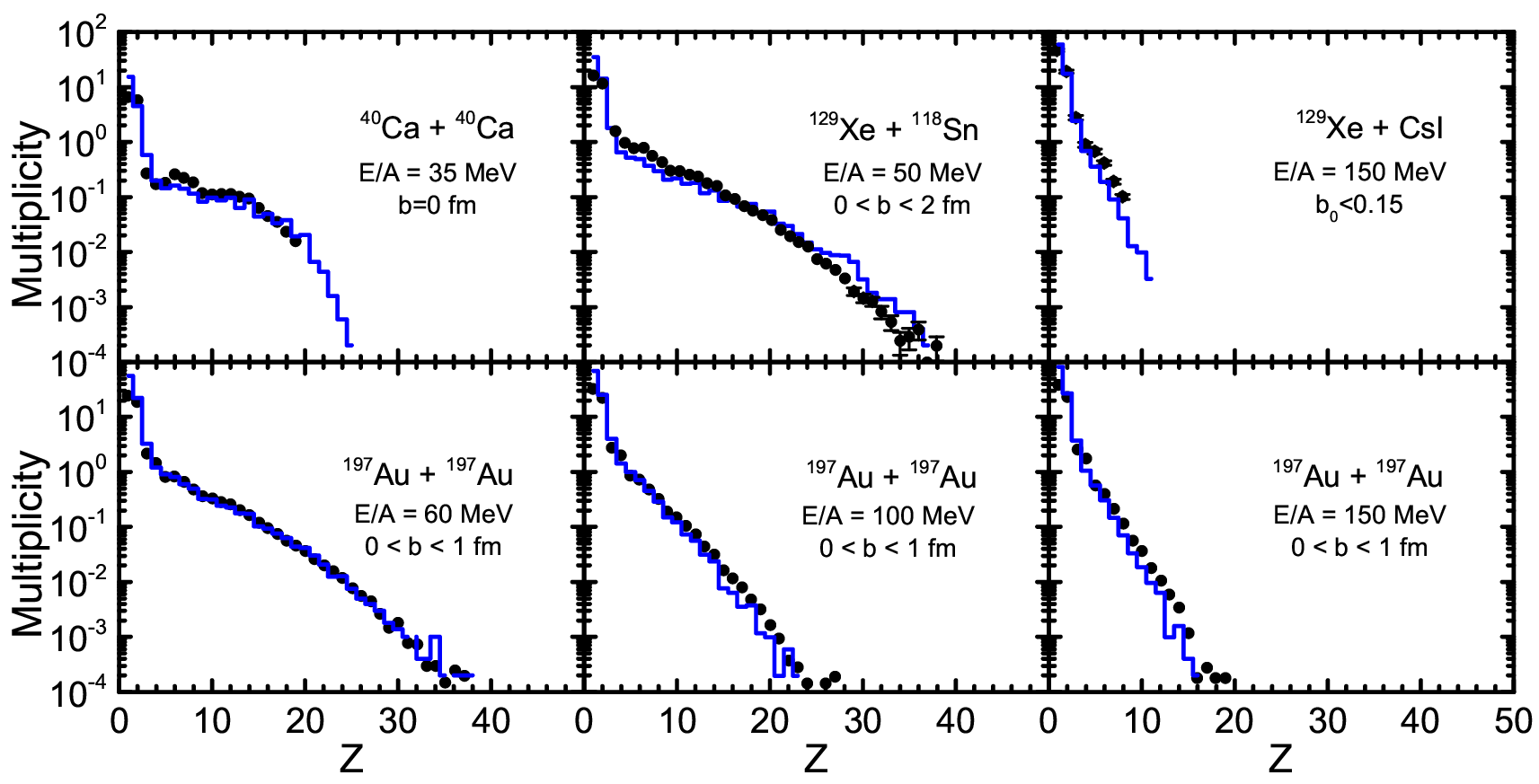}
	\caption{The fragment charge distributions in central HIC are plotted in solid lines for various reaction systems in comparison with the experimental data(solid circles). The experimental results are respectively taken from Ref.\cite{Ha94} for $^{40}$Ca+$^{40}$Ca, the INDRA data\cite{Hu03} for $^{129}$Xe+$^{118}$Sn, the FOPI data\cite{Re10} for $^{129}$Xe+CsI, and the INDRA data\cite{Zb07} for all the $^{197}$Au+$^{197}$Au collisions in the lower panels. For the model, only the raw(unfiltered) results with centralities selected by an impact parameter cut are given.}
\end{figure*}
%%%%%%%%%%%%%%%%%%%%%%%%%%%%%%%%%%%%%%%%%%%%%%%%%%%%%%%%%%%%%%%

In Fig. 3, we show the charge distributions of fragments reproduced by our model in comparison with the experimental data, in central HIC for typical reaction systems. Here a single set of MST parameters $r_{0}=3.5$ fm and $p_{0}=200$ MeV/c is adopted. Roughly speaking, our model gives a reasonable reproduction of the experimental charge spectra for $Z\geq2$. At higher energies, our model seems to underestimate the data for fragments of large charge number, as also seen in the QMD results in Ref.\cite{Zb07}. This is due the the scheme of centrality selection we adopted here and the fact that the results presented here are raw and unfiltered. But the situation for $^{197}$Au+$^{197}$Au can be very much improved by applying a different scheme of centrality selection and the INDRA filter\cite{Zb07}. For $^{129}$Xe+$^{118}$Sn@50A MeV, the yields of Intermediate-Mass-Fragment (IMF) which have more than one nucleon occupying a spin-isospin state, are underestimated due to a stronger phase-space constraint on the boundary regions, the breeding place not only for clusters but also for IMF. In the meantime, the yields of heavy fragments are overestimated due to the lack of fragmentation, which can be explained by the incompleteness of our description of the fermionic nature of nucleons. Another interesting result is that, in our model, the yield of fragments of $Z=1$ are very much lower than without cluster correlation\cite{Zb07} for $^{197}$Au+$^{197}$Au, which is a sign that the missing protons appear in gas-phase clusters. This is most apparent in $^{129}$Xe+CsI@150A MeV where the yield of fragments of $Z=1$ is close to the experimental value. In this particular reaction system, the production of very large fragments is negligible and the produced fireball ends up more gas-like, which suppresses the spurious emission of nucleons from hot big fragments due to the classical nature of the description by QMD. With the above observations, we can hitherto conclude that our model succeeds in a reasonable and unified description of both the yields of light clusters and heavier fragments, which is a promising feature once the model is refined and extended to higher energies to describe clusters in spectator fragmentation concerning strangeness, a drawback common to the BUU-type model.
%%%%%%%%%%%%%%%%%%%%%%%%%%%%%%%%%%%% figure 4 %%%%%%%%%%%%%%%%%%%%%%
\begin{figure}
	\includegraphics[width=8.5 cm]{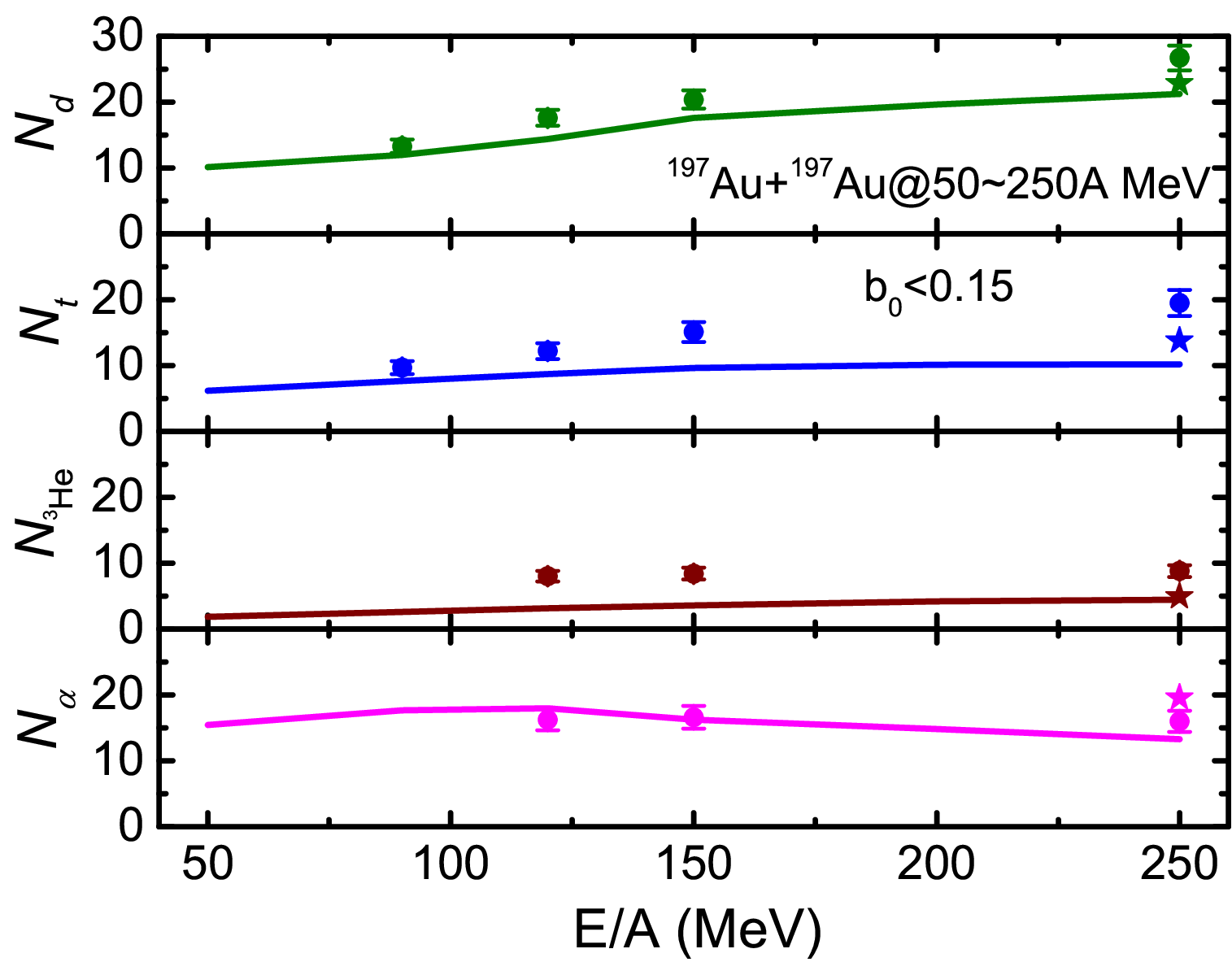}
	\caption{The dependence of the yields of light clusters including the deuteron(olive), triton(blue), $^{3}$He(wine) and the $\alpha$ particle(magenta) on the incident energy is plotted in solid lines for central $^{197}$Au+$^{197}$Au collisions, in comparison with the FOPI data\cite{Re10}. Here the results of AMD-cluster\cite{On16} are also displayed in solid stars alongside for comparison.}
\end{figure}
%%%%%%%%%%%%%%%%%%%%%%%%%%%%%%%%%%%%%%%%%%%%%%%%%%%%%%%%%%%%%%%

Finally, in Fig. 4, the dependence of the yields of light clusters on the incident energies is shown for central HIC with the reaction system $^{197}$Au+$^{197}$Au, in comparison with available experimental data and the results of AMD-cluster\cite{On16}. Firstly, we see that at the highest energy 250A MeV, our result is very similar to that of AMD-cluster, with the yields of deuterons, tritons and $^{3}$He underestimated, which is within our expectation since both models employ the same prescription in the treatment of cluster correlation. Since, this treatment has very few tunable parameters, we are confident that combined with a more flexible parametrization such as that in Ref.\cite{Wa23}, a satisfactory description of light cluster production over the whole energy range from several tens of MeV to 1 GeV will be obtained in a straight forward way. Finally, our model correctly describes the increasing trend of the yield of deuterons against the incident energy, and the experimental fact that the yield of $\alpha$ is almost a plateau over the shown incident energy range.

%Conclusions
In conclusion, by implementing cluster correlation, a popular microscopic transport approach QMD is for the first time shown to be as well powerful in a unified description of the production of light clusters and heavier fragments in heavy-ion collisions, which may open the perspective for various future investigations. A method of continuous phase-space constraint based on the technique of frictional cooling is devised and acts like a pauli potential to improve the phase-space distribution of nucleons, and facilitate the formation and emission of clusters through pauli repulsion during the course of heavy-ion collisions. The binding energy of clusters is also considered and is important for the description of cluster production. A reasonable reproduction of both the experimental light cluster multiplicities including deuterons, tritons, $^{3}$He and $\alpha$, and the experimental charge distributions of heavier fragments is achieved. The isospin asymmetries of the reaction systems are well reflected on the ratio triton/$^{3}$He, and as far as light cluster production is concerned, the results are insensitive to the time cut to switch from QMD to the statistical afterburner. The trends of yield of different kinds of light clusters against incident energy is correctly reproduced. Imperfect it may be, this work is a breakthrough for this transport approach and it paves the first successful path towards the study of light cluster production within this transport approach in multi-fragmentation reactions at intermediate energies. This makes accessible the perspective for various promising future extensions, for example, hypercluster formation in spectator fragmentation reactions, in which this approach has special advantages.

\section{Acknowledgements}
The authors are indebted to Professor Akira Ono for helpful discussions on the details of the treatments and the relevant physics. This work was supported by the National Natural Science Foundation of China (Projects No. 12175072 and No. 12311540139) and the Talent Program of South China University of Technology (Projects No. 20210115).

\end{document}